\documentstyle[12pt]{article}
\def\thebibliography#1{\section*{REFERENCES}\list
  {[\arabic{enumi}]}{\settowidth\labelwidth{#1}\leftmargin\labelwidth
    \advance\leftmargin\labelsep
    \usecounter{enumi}}
    \def\newblock{\hskip .11em plus .33em minus .07em}
    \sloppy\clubpenalty4000\widowpenalty4000
    \sfcode`\.=1000\relax}

\let\Large=\large

\def\op#1{\mathop{\fam0 #1}\limits}

\newcommand{\nm}[1]{\mid {#1}\mid}

\newcommand{\beq}{\begin{equation}}
\newcommand{\eeq}{\end{equation}}
\newcommand{\ben}{\begin{eqnarray}}
\newcommand{\een}{\end{eqnarray}}
\newcommand{\be}{\begin{eqnarray*}}
\newcommand{\ee}{\end{eqnarray*}}
\newcommand{\bea}{\begin{eqalph}}
\newcommand{\eea}{\end{eqalph}}
\newcommand{\bR}{{\bf R}}
\newcommand{\al}{\alpha}
\newcommand{\bt}{\beta}
\newcommand{\la}{\lambda}
\newcommand{\f}{\phi}

\newcommand{\m}{\mu}

\newcommand{\g}{\gamma}
\newcommand{\G}{\Gamma}

\newcommand{\si}{\sigma}

\newcommand{\ol}{\overline}
\newcommand{\dr}{\partial}

\newcommand{\ar}{\op\longrightarrow}
\newcommand{\ot}{\otimes}

\newcounter{eqalph}
\newcounter{equationa}
\newcounter{theorem}
\newcounter{proposition}
\newcounter{lemma}
\newcounter{corollary}
\newcounter{definition}

\newenvironment{eqalph}{\stepcounter{equation}
\setcounter{equationa}{\value{equation}}
\setcounter{equation}{0}

\begin{eqnarray}}{\end{eqnarray}\setcounter{equation}{\value{equationa}}}

\def\thedefinition{\arabic{definition}}

\newenvironment{rem}{\medskip\noindent{\it Remark.}}{\medskip}

\newenvironment{prop}{\refstepcounter{definition} 
\bigskip\noindent{\it Proposition \thedefinition.}}{\medskip}
\newenvironment{lem}{\refstepcounter{definition} 
\bigskip\noindent{\it Lemma \thedefinition.}}{\medskip}

\begin{document}

\hbox{}

{\parindent=0pt 

{ \Large \bf Nonrelativistic Geodesic Motion}
\bigskip

{\bf Luigi Mangiarotti\footnote{Department of Mathematics
and Physics, University of Camerino, 62032 Camerino (MC), Italy; e-mail
mangiaro@camserv.unicam.it} and Gennadi  Sardanashvily\footnote{Department of
Theoretical Physics, Physics Faculty, Moscow State University, 117234 Moscow,
Russia; e-mail: sard@grav.phys.msu.su}}
\bigskip

We show that any second order dynamic equation on a
configuration space $X\to\bR$ of nonrelativistic time-dependent mechanics  can
be seen as a geodesic equation with respect to some (nonlinear) connection on
the tangent bundle $TX\to X$ of relativistic velocities. We compare
relativistic and nonrelativistic geodesic equations, and study the Jacobi
vector fields along nonrelativistic geodesics. }

\section{INTRODUCTION}

To provide a geometric formulation of nonrelativistic
mechanics, one usually try to introduce a metric on a configuration space.
Following Cartan's idea, we  show that any second order dynamic equation of
nonrelativistic mechanics is equivalent to a particular geodesic equation on a
phase space of relativistic 4-velocities. The key point is that relativistic
and nonrelativistic geodesic equations  are defined on different subspaces of
the same 4-velocity phase space.  One can perform a relativization of
nonrelativistic dynamic equations by means of their extension onto the
relativistic subspace of the 4-velocity phase space. However, such an
extension fails to be unique. In Section 4, we will consider the most
important examples. Treating nonrelativistic dynamic equations as the geodesic
ones, we can study them by means of the well-known differential geometric
methods. In particular, Jacobi vector fields along nonrelativistic geodesics
can be introduced in a natural way, and conjugate points of these geodesics can
be investigated (see Section 5).

Let $X$ be a 4-dimensional world manifold of a relativistic theory,
coordinated by
$(x^\la)$. Then the tangent bundle $TX$ of $X$ plays the role
of a 4-velocity phase space.
By a relativistic equation of motion usually is meant
a geodesic equation
\beq
\ddot x^\m= K_\la^\m(x^\nu,\dot
x^\nu) \dot x^\la
\label{cqg2}
\eeq
with respect to a (nonlinear) connection 
\beq
K=dx^\la\ot(\dr_\la +K^\m_\la\dot\dr_\m) \label{cqg3}
\eeq
on the tangent bundle $TX\to X$. 
It is supposed additionally that there is a pseudo-Riemannian metric
$g$ of signature $(+,---)$ on $X$, and that a geodesic vector field
does not leave the subbundle of
relativistic hyperboloids 
\beq
W_g=\{\dot x^\la\in TX\, \mid \,\,g_{\la\m} \dot x^\la\dot x^\m=1\}
\label{cqg1}
\eeq
in $TX$. It suffices to require that the condition
\beq
(\dr_\la g_{\m\nu}\dot x^\m + 2g_{\m\nu}K^\m_\la)\dot x^\la \dot x^\nu =0
\label{cqg4}
\eeq
holds for all tangent vectors which belong to $W_g$ (\ref{cqg1}). 
 Of course, the
Levi--Civita connection 
$\{_\la{}^\m{}_\nu\}$ of the metric $g$ fulfills the
condition (\ref{cqg4}). Any connection
$K$ on the tangent bundle $TX\to X$ can be written as
\be
K^\m_\la = \{_\la{}^\m{}_\nu\}\dot x^\nu + \si^\m_\la(x^\la,\dot x^\la),
\ee
where the soldering form $\si=\si^\m_\la dx^\la\ot\dot\dr_\la$
plays the role of an external force. Then the condition (\ref{cqg4}) takes
the form
\beq
g_{\m\nu}\si^\m_\la\dot x^\la \dot x^\nu=0. \label{cqg46}
\eeq

Let now a world manifold $X$ admit a projection $X\to \bR$, where $\bR$ is
a time axis. One can think of the bundle $X\to\bR$ as being the configuration
space of nonrelativistic mechanics (Massa and Pagani,
1994; Mangiarotti and Sardanashvily, 1998; Sardanashvily, 1998). It is
provided with the adapted bundle coordinates
$(x^0,x^i)$, where the transition functions of the temporal one are $x'^0=x^0
+$const. The corresponding velocity phase space is the first
order jet manifold
$J^1X$ of $X\to \bR$, coordinated by $(x^\la, x^i_0)$. There is the canonical
imbedding of 
$J^1X$ onto the affine subbundle of the tangent
bundle $TX$ (see (\ref{z260}) below), given by the coordinate conditions 
\beq
\dot x^0=1, \qquad \dot x^i=x^i_0. \label{cqg7}
\eeq
Then one can regard (\ref{cqg7}) as the 4-velocities of a
nonrelativistic system. The relation (\ref{cqg7}) differs from the
familiar relation between 4- and 3-velocities of a relativistic
system. It follows that the 4-velocities of
relativistic and nonrelativistic systems occupy different subbundles of the
tangent bundle
$TX$. The key point of our consideration is the following.

\begin{prop} \label{c1} Let $J^2X$ be the second order jet manifold
of $X\to\bR$, coordinated by $(x^\la,x^i_0,x^i_{00})$.  Any second order
dynamic equation 
\beq
x^i_{00} =\xi^i(x^0,x^j,x_0^j) \label{cqg5}
\eeq
of nonrelativistic mechanics on $X\to\bR$ is equivalent to the geodesic
equation 
\ben
 && \ddot x^0 =0, \qquad \dot x^0=1, \nonumber\\
 && \ddot x^i= \ol K^i_0 \dot x^0 + \ol K^i_j\dot x^j
\label{cqg11}
\een
with respect to a connection
$\ol K$ on $TX\to X$ which fulfills the conditions
\beq
\ol K^0_\la =0, \qquad
 \xi^i= \ol K^i_0 + x^j_0\ol K^i_j\mid_{\dot x^0=1,\dot x^i=x^i_0}.
\label{cqg9}
\eeq 
\end{prop}

Thus, we observe that both relativistic and nonrelativistic equations of
motion can be seen as the geodesic equations on the same tangent bundle
$TX$.  The difference between them lies in the fact that their solutions
live in the different subbundles (\ref{cqg1}) and (\ref{cqg7}) of $TX$.  
At the same time, relativistic equations, expressed into the  3-velocities
$x^i_0=\dot x^i/\dot x^0$, tend exactly to the
nonrelativistic  equations on the subbundle (\ref{cqg7}) when $\dot x^0\to
1$, $g_{00}\to 1$, i.e., only when nonrelativistic mechanics and the
nonrelativistic approximation of a relativistic theory coincide.

\section{GEOMETRY OF NONRELATIVISTIC MECHANICS}

This Section is devoted to the proof of Proposition 1.
Let a fibre bundle $X\to\bR$, coordinated by $(x^0,x^i)$, be a configuration
space of nonrelativistic mechanics. Its velocity phase space $J^1X$ 
 is provided with the adapted coordinates
$(x^0,x^i,x^i_0)$. Recall that $J^1X$ comprises the equivalence classes
$j^1_{x^0}c$ of sections of
$X\to\bR$ which are identified by their values  $c^i(x^0)$ and the values of
their derivatives
$\dr_0c^i(x^0)$ at points $x^0\in\bR$, i.e., $x^i_0(j^1c)= \dr_0c^i(x^0)$.
There is the canonical imbedding 
\beq
\la: J^1X\hookrightarrow TX, \qquad \la=\dr_0 +x^i_0\dr_i,\label{z260}
\eeq
over $X$. 
From now on, we will 
identify $J^1X$ with its image in $TX$. 
It is an
affine bundle modelled over the vertical tangent bundle $VX$ of
$X\to\bR$. 

In
particular, every connection on a bundle $X\to\bR$ is given by the nowhere 
vanishing vector field 
\beq
\G:X\to J^1X\subset TX, \qquad \G=\dr_0 +\G^i\dr_i, \label{1005}
\eeq
on $X$. It can be treated as a reference frame in nonrelativistic mechanics.
Every connection $\G^i$ (\ref{1005}) defines an atlas of local constant
trivializations of the bundle $X\to\bR$ and the associated coordinates
$(x^0,x^i)$ on $X$ such that the transition functions
$x^i\to x'^i$ are independent of
$x^0$,  and {\it vice versa} (Mangiarotti and Sardanashvily, 1998;
Mangiarotti {\it et al.}, 1999). We find
$\G^i=0$ with respect to this coordinate atlas, also called a reference frame.
In particular, there is one-to-one correspondence between the complete
connections
$\G$ (\ref{1005}) and the trivializations $X\cong
\bR\times M$ of the configuration
bundle $X$. 

A nonrelativistic second order dynamic equation on a configuration bundle
$X\to\bR$ is defined as the geodesic equation 
\be
x^i_{00}=\xi^i(x^\m,x^j_0) 
\ee
for a holonomic connection
\beq
\xi=\dr_0 + x^i_0\dr_i + \xi^i(x^\m,x^i_0) \dr_i^0 \label{a1.30}
\eeq
on the jet bundle $J^1X\to\bR$, which takes its values into
the second order jet manifold 
$J^2X$. It has the transformation law
\be
\xi'^i=(\xi^j\dr_j + x^j_0x^k_0\dr_j\dr_k
+2x^j_0\dr_j\dr_0 +\dr_0^2)x'^i.
\ee

Let us consider the relationship
between the holonomic connections $\xi$ (\ref{a1.30}) on the jet bundle
$J^1X\to\bR$ and the connections 
\beq
 \g=dx^\la\ot (\dr_\la + \g^i_\la \dr_i^0)
\label{a1.38} 
\eeq 
on the affine jet bundle $J^1X\to X$. The connections $\g$ have the
transformation law
\beq
\g'^i_\la = (\dr_jx'^i\g^j_\m
+\dr_\m x'^i_0)\frac{\dr x^\m}{\dr x'^\la}. \label{m175}
\eeq

\begin{prop}\label{gena51}
Any connection $\g$ (\ref{a1.38}) on the affine jet bundle $J^1X\to X$ defines
the holonomic connection 
\beq
\xi_\g = \dr_0 + x^i_0\dr_i +(\g^i_0 +x^j_0\g^i_j)\dr_i^0
\label{z281}
\eeq
on the jet bundle $J^1X\to\bR$ (De Le\'on and Rodrigues, 1989; Mangiarotti and
Sardanashvily, 1998; Mangiarotti {\it et al.}, 1999).
\end{prop}

It follows that every connection $\g$ (\ref{a1.38}) on the affine jet bundle 
$J^1X\to X$ yields the dynamic equation
\beq
x^i_{00}=\g^i_0 +x^j_0\g^i_j \label{z287}
\eeq
on the configuration space $X$. 
Of course, different dynamic connections may lead to the same dynamic
equation (\ref{z287}). The converse assertion is the following
(Crampin {\it et al.}, 1996; Mangiarotti and
Sardanashvily, 1998; Mangiarotti {\it et al.}, 1999).

\begin{prop}\label{gena52}
Any holonomic connection $\xi$ (\ref{a1.30}) on the jet bundle
$J^1X\to \bR$ defines a connection 
\beq
\g =dx^0\ot[\dr_0+(\xi^i-\frac12 x^j_t\dr_j^t\xi^i)\dr_i^0] +
dx^j\ot[\dr_j +\frac12\dr_j^0\xi^i \dr_i^0]
\label{z286}
\eeq
on the affine
jet bundle $J^1X\to X$.
\end{prop}

The connection $\g$ (\ref{z286}),
associated with a dynamic equation,  possesses the property
\be
\g^k_i = \dr_i^0\g^k_0 +  x^j_0\dr_i^0\g^k_j,
\ee
which implies the relation $\dr_j^0\g^k_i = \dr_i^0\g^k_j$. Such
a connection $\g$ is
called symmetric.

Let $\g$ be a connection (\ref{a1.38}) and $\xi_\g$ the 
corresponding dynamic equation (\ref{z281}). Then 
the connection (\ref{z286}), associated with $\xi_\g$, takes the form
\be
\g_{\xi_\g}{}^k_i = \frac{1}{2}
(\g^k_i + \dr_i^0\g^k_0 + x^j_0\dr_i^0\g^k_j),
\qquad \g_{\xi_\g}{}^k_0 = \xi^k - x^i_0\g_{\xi_\g}{}^k_i. 
\ee
It is readily observed that $\g = \g_{\xi_\g}$ if and only if $\g$ is
symmetric.
 
Now let us prove Proposition \ref{c1}.
We start from the relation between the connections $\g$ on the affine
jet bundle $J^1X\to X$ and the connections
$K$ (\ref{cqg3})
on the tangent bundle $TX\to X$ of the configuration space $X$.  Let us
consider the diagram
\beq
\begin{array}{rcccl}
& J^1_XJ^1X & \ar^{J^1\la} & J^1_XTX & \\
_\g &  \put(0,-10){\vector(0,1){20}} & &  \put(0,-10){\vector(0,1){20}}
& _K\\
& J^1X &\ar^\la & TX &
\end{array} \label{z291}
\eeq
where $J^1_XJ^1X$ is the first order jet manifold of the affine jet bundle
$J^1X\to X$ with coordinates $(x^\la,x^i_0, x^i_{\m 0})$ and 
$J^1_XTX$ is the first order jet manifold of the tangent bundle $TX\to X$,
coordinated by
$(x^\la,\dot x^\la,\dot x^\la_\m)$. The jet prolongation over $X$ of the
canonical imbedding $\la$ (\ref{z260}) reads 
\be
J^1\la: (x^\la,x^i_0, x^i_{\m 0}) \mapsto 
(x^\la,\dot x^0=1,\dot x^i=x^i_0, \dot x_\m^0=0,
\dot x^i_\m=x^i_{\m 0}).
\ee
We have
\be
&& J^1\la\circ \g: (x^\la,x^i_0) \mapsto 
(x^\la,\dot x^0=1,\dot x^i=x^i_0, \dot x_\m^0=0,
\dot x^i_\m=\g^i_\m ),\\
&& K\circ \la: (x^\la,x^i_0) \mapsto 
(x^\la,\dot x^0=1,\dot x^i=x^i_0, \dot x^0_\m=K_\m^0,
\dot x^i_\m=K^i_\m).
\ee
It follows that the diagram (\ref{z291}) can be commutative only
if the components $K^0_\m$ of the connection $K$ on $TX\to
X$ vanish. 
Since the transition functions $x^0\to x'^0$ are independent of
$x^i$, a connection $K$ with the components $K^0_\m=0$ can exist on the tangent
bundle
$TX\to X$. In particular, let $(x^0,x^i)$ be a reference frame. Given
an arbitrary connection $K$ (\ref{cqg3}) on $TX\to X$, one can put 
$K^0_\m=0$ in order to obtain a desired connection
\beq
\ol K= dx^\la\ot (\dr_\la +K^i_\la\dot\dr_i), 
\label{z292}
\eeq
obeying the transformation law
\beq
{K'}_\la^i=(\dr_j x'^i K^j_\m + \dr_\m\dot x'^i)
\frac{\dr x^\m}{\dr x'^\la}.  \label{z293}
\eeq
Now the diagram (\ref{z291}) becomes commutative if the connections
$\g$ and $\ol K$ fulfill the relation
\beq
\g^i_\m=K^i_\m(x^\la,\dot x^0=1, \dot x^i=x^i_0).
\label{z294}
\eeq
It is easily seen that this relation holds globally because the
substitution of $\dot x^i=x^i_0$ into (\ref{z293}) restates the
transformation law (\ref{m175}).  In accordance with the relation
(\ref{z294}), a desired connection $\ol K$ is an extension  of the 
section
$J^1\la\circ \g$ of the affine bundle $J^1_XTX\to TX$ over the closed
submanifold $J^1X\subset TX$ to a global section. Such an extension
always exists, but it is not unique. Thus, it is stated the following.

\begin{prop}
In accordance with the relation (\ref{z294}), 
every dynamic equation on the configuration space $X$ can be written in the
form 
\beq
x^i_{00} = K^i_0\circ\la +x^j_0 K^i_j\circ\la, \label{gm340}
\eeq
where $\ol K$ is a connection (\ref{z292}) on the tangent bundle $TX\to X$.
\end{prop}

Let us consider the geodesic equation (\ref{cqg11}) on $TX$ with respect to
the connection $\ol K$. Its solution is a geodesic curve $c(t)$ also
satisfying the dynamic equation (\ref{cqg5}), and {\it vice versa}. It states
Proposition 1.

The above proof also leads to the following converse of Proposition one.

\begin{prop} \label{c1'}
Given a reference frame,  any connection $K$
(\ref{cqg3}) on the tangent bundle $TX\to X$ defines a connection 
$\g$ on the affine jet bundle $J^1X\to X$ and the dynamic equation
(\ref{gm340}) on the configuration space $X$.
\end{prop}

\begin{rem}
Note that any second order dynamic equation on $Q\to\bR$ also defines a
linear connection on the tangent bundle $TJ^1Q\to J^1Q$ (Massa and Pagani,
1994; Crampin {\it et al.}, 1996; Mangiarotti and Sardanashvily, 1998). A
conservative second order dynamic equation on a manifold $Z$ also defines a
connection on $TZ\to Z$, but it is a geodesic equation with respect to this
connection if and only if this connection is a spray (Marmo {\it et al.},
1990; Morandi {\it et al.}, 1990; Mangiarotti and Sardanashvily, 1998).
\end{rem}

\section{QUADRATIC DYNAMIC EQUATIONS}

From the physical viewpoint, the most interesting dynamic equations are the
quadratic ones, i.e.,
\beq
\xi^i = a^i_{jk}(x^\m)x^j_0 x^k_0 + b^i_j(x^\m)x^j_0 + f^i(x^\m).
\label{cqg100}
\eeq
This property is coordinate-independent due to the affine transformation law
of coordinates $x^i_0$.
Then, it is readily observed that the corresponding connection $\g$
(\ref{z286}) is affine:
\be
 \g=dx^\la\ot [\dr_\la + (\g^i_{\la 0}(x^\nu)+ \g^i_{\la
j}(x^\nu)x^j_0)\dr_i^0],
\ee
 and {\it vice versa}.
This connection is symmetric if and only if $\g^i_{\la \m}=\g^i_{\m\la}$. 

\begin{lem}\label{aff}
There is one-to-one correspondence between the affine connections $\g$ on
the affine jet bundle $J^1X\to X$ and the linear connections $\ol K$
(\ref{z292}) on the tangent bundle $TX\to X$. This correspondence is
given by the relation (\ref{z294}) which takes the form
\be
\g^i_\m=\g^i_{\m 0} + \g^i_{\m j}x^j_0, \qquad  \g^i_{\m\la}= K_\m{}^i{}_\la. 
\ee
\end{lem}

In particular, if an affine connection $\g$ is symmetric, so is the
corresponding linear connection $\ol K$. Then we come to the following
corollaries of Propositions \ref{c1} and \ref{c1'}.

\begin{prop}\label{c2} 
Any quadratic dynamic equation 
\beq
x^i_{00}= a^i_{jk}(x^\m)x^j_0 x^k_0 + b^i_j(x^\m)x^j_0 + f^i(x^\m)
\label{cqg100'}
\eeq
is equivalent to the geodesic
equation 
\ben
&& \ddot x^0= 0, \qquad \dot x^0=1,\nonumber\\
&& \ddot x^i= 
a^i_{jk}(x^\m)\dot x^i \dot x^j + b^i_j(x^\m)\dot x^j\dot x^0 +
f^i(x^\m) \dot x^0\dot x^0 \label{cqg17}
\een
for the symmetric linear connection 
\be
\ol K=dx^\la\ot(\dr_\la + K_\la{}^\m{}_\nu(x^\al)\dot x^\nu\dot\dr_\m)
\ee
on $TX\to X$,
given by the components
\beq
K_\la{}^0{}_\nu=0, \quad K_0{}^i{}_0= f^i, \quad
K_0{}^i{}_j=K_j{}^i{}_0=\frac12 b^i_j,
\quad K_j{}^i{}_k= a^i_{jk}. \label{cqg101}
\eeq
\end{prop}

\begin{prop}\label{aff1}
Conversely, any linear connection $K$  on the tangent bundle
$TX\to X$ defines the quadratic dynamic equation
\be
x^i_{00}= K_0{}^i{}_0 +(K_0{}^i{}_j +K_j{}^i{}_0)x^j_0 + K_j{}^i{}_kx^j_0,
x^k_0,
\ee
written with respect to a given reference frame $(x^0,x^i)$.
\end{prop}

The geodesic equation
(\ref{cqg17}) however is not unique for the dynamic equation (\ref{cqg100'}).

\begin{prop} \label{jp40} 
Any quadratic dynamic equation (\ref{cqg100'}), being 
equivalent to the geodesic equation with respect to the linear connection
$\ol K$ (\ref{cqg101}), is also equivalent to the geodesic equation with
respect to an affine connection $K'$ on $TX\to X$ which differs from $\ol K$
(\ref{cqg101}) in a soldering form $\si$ on $TX\to X$ with the components
\be
\si^0_\la= 0, \qquad \si^i_k= h^i_k+ (s-1) h^i_k\dot x^0, \qquad \si^i_0=
-s h^i_k\dot x^k -h^i_0\dot x^0 + h^i_0,
\ee
where $s$ and $h^i_\la$ are local functions on $X$.
\end{prop} 

In particular, it follows that, if there is no
topological obstruction and the Minkowski metric $\eta$ on $TX$ exists, a
nonrelativistic dynamic equation 
\beq
x^i_{00}= b^i_j(x^\m)x^j_0 + f^i(x^\m)
\label{cqg18}
\eeq
gives rise to the geodesic equation
\ben
&& \ddot x^0= 0, \qquad \dot x^0=1, \nonumber\\
&& \ddot x^i=  b^i_j(x^\m)\dot x^j +
f^i(x^\m) \dot x^0. \label{cqg19}
\een

The above-mentioned ambiguity often occurs.
The nonrelativistic dynamic equations (\ref{cqg18}) can be
represented as both the geodesic equation (\ref{cqg19}) and the one
(\ref{cqg17}), where $a=0$. The first is the case for
external forces, e.g., an electromagnetic theory, while the latter is that for
a gravitation theory.

\section{EXAMPLES}

In order to compare relativistic and nonrelativistic dynamics, one should
consider a pseudo-Riemannian metric on $TX$, compatible with the fibration
$X\to\bR$. Note that $\bR$ is a time of nonrelativistic mechanics. It
is one for all nonrelativistic observers. In the framework of a relativistic
theory, this time can be seen as a cosmological time.  Given  a fibration
$X\to\bR$, a pseudo-Riemannian metric on the tangent bundle $TX$ is said to be
admissible if it is defined by a pair $(g^R,\G)$ of a Riemannian metric $g^R$
on $X$ and a nonrelativistic reference frame $\G$ (\ref{1005}), i.e.,
\beq
g=\frac{2\G\ot\G}{\nm\G^2} - g^R, \quad
\nm\G^2=g^R_{\m\nu}\G^\m\G^\nu= g_{\m\nu}\G^\m\G^\nu, \label{cqg32}
\eeq
in accordance
with the well-known theorem (Hawking and Ellis, 1973).
The vector field $\G$ is time-like relative to the
pseudo-Riemannian metric $g$ (\ref{cqg32}), but not with respect to other
admissible pseudo-Riemannian metrics in general.

As we have shown above, given  a reference frame $(x^0,x^i)$, any connection
$K(x^\la,\dot x^\la)$ (\ref{cqg3}) on the tangent bundle
$TX\to X$ defines the connection
$\ol K$ on $TX\to X$ with the components
\beq
\ol K^0_\la =0, \qquad \ol K^i_\la=K^i_\la. \label{cqg102}
\eeq
It follows that, given
a fibration $X\to\bR$, every relativistic equation of motion (\ref{cqg2})
yields the geodesic equation (\ref{cqg11}) and, consequently, has the
counterpart  
\be
x^i_{00}=K^i_0(x^\la,1,x^k_0) +K^i_j(x^\la,1,x^k_0)x^j_0 
\ee
(\ref{cqg5}) in nonrelativistic mechanics. 
Note that, written with respect to a
reference frame $(x^0,x^i)$, the connection $\ol K$ (\ref{cqg9}) and the
corresponding geodesic equation (\ref{cqg11}) are well defined relative to
any coordinates on
$X$, while the dynamic equation (\ref{cqg2}) is done relative to arbitrary
coordinates on $X$, compatible with the fibration $X\to\bR$. The key point is
that, for another reference frame $(x^0,x'^i)$ with
time-dependent transition functions $x^i\to x'^i$, the same connection $K$
(\ref{cqg3}) on $TX$ sets another connection $\ol K'$  on $TX\to X$ with the
components
\be
K'^0_\la=0, \qquad K'^i_\la = \left(\frac{\dr x'^i}{\dr x^j}K^j_\m +\frac{\dr
x'^i}{\dr x^\m}\right)\frac{\dr x^\m}{\dr x'^\la} +
\frac{\dr x'^i}{\dr x^0} \frac{\dr x^\m}{\dr x'^\la}K^0_\m,
\ee 
while the
connection
$\ol K$ (\ref{cqg102}) has the components
\be
K'^0_\la=0, \qquad K'^i_\la = \left(\frac{\dr x'^i}{\dr x^j}K^j_\m +\frac{\dr
x'^i}{\dr x^\m}\right)\frac{\dr x^\m}{\dr x'^\la},
\ee
relative to the same reference frame.
This illustrates the obvious fact that a nonrelativistic approximation is not
relativistic invariant (see, e.g. (L\'evy--Leblond, 1967)).

The converse procedure is more intricate. At first, a nonrelativistic
dynamic equation (\ref{cqg5}) is brought into the geodesic equation
(\ref{cqg11}) with respect to the connection $\ol K$ (\ref{cqg9}). A solution
is not unique in general. Then, one should find a pair $(g,K)$  of a
pseudo-Riemannian metric
$g$ and a connection $K$ on $TX\to X$ such $K^i_\la=\ol K^i_\la$ and the
condition (\ref{cqg4}) is fulfilled.

Given a coordinate systems $(x^0,x^i)$, compatible with the fibration $X\to
\bR$, let us consider a nondegenerate quadratic Lagrangian 
\beq
L=\frac12m_{ij}(x^\m) x^i_0 x^j_0 + k_i(x^\m) x^i_0  +
f(x^\m), \label{cqg20}
\eeq
where $m_{ij}$ is a Riemannian mass tensor. Similarly to
Lemma
\ref{aff}, one can show that any quadratic polynomial on $J^1X\subset TX$ is
extended to a bilinear form on $TX$. Then the Lagrangian $L$
(\ref{cqg20}) can be written as 
\beq
L=-\frac12g_{\al\m}x^\al_0 x^\m_0, \qquad x^0_0=1, \label{cqg40}
\eeq
where $g$ is the metric
\beq
g_{00}=-2f, \qquad g_{0i}=-k_i, \qquad g_{ij}=-m_{ij} \label{cqg21}
\eeq
on $X$. The corresponding Lagrange equation takes the form
\beq
x^i_{00}=-(m^{-1})^{ik}\{_{\la k\nu}\}x^\la_0x^\nu_0, \qquad x^0_0=1,
\label{cqg35}
\eeq
where 
\be
\{_{\la\m\nu}\} =-\frac12(\dr_\la g_{\m\nu} +\dr_\nu
g_{\m\la} - \dr_\m g_{\la\nu})
\ee
 are the Christoffel symbols of the metric (\ref{cqg21}). Let us assume that
this metric is nondegenerate. By virtue of
Proposition \ref{c2}, the dynamic equation (\ref{cqg35}) can be brought into
the geodesic equation (\ref{cqg17}) on $TX$ which reads
\ben
&& \ddot x^0 = 0, \qquad \dot x^0=1, \nonumber \\
&&\ddot x^i = (\{_\la{}^i{}_\nu\} -
\frac{g^{k0}}{g^{00}}\{_\la{}^0{}_\nu\})\dot x^\la\dot x^\nu. \label{cqg47}
\een

Let us now bring the Lagrangian (\ref{cqg20}) into the form
\beq
L=\frac12m_{ij}(x^\m)( x^i_0-\G^i)(x^j_0-\G^j) +
f'(x^\m), \label{cqg48}
\eeq
where $\G$ is a Lagrangian connection on $X\to\bR$. This connection $\G$
defines an atlas of local constant trivializations of the bundle $X\to\bR$
and the corresponding coordinates $(x^0,\ol x^i)$ on $X$ such that the
transition functions $\ol x^i\to \ol x'^i$ are independent of $x^0$, and
$\G^i=0$ with respect to $(x^0,\ol x^i)$. In this
coordinates, the Lagrangian $L$ (\ref{cqg48}) reads
\be
L=\frac12\ol m_{ij}\ol x^i_0 \ol x^j_0 +
f'(x^\m). 
\ee
One can think of its first term as the kinetic energy of
a nonrelativistic system with the mass tensor $\ol m_{ij}$ relative to the
reference frame
$\G$, while $(-f')$ is a potential.  Let us assume that
$f'$ is a nowhere vanishing function on
$X$, i.e., the metric (\ref{cqg21}) is nondegenerate. Then the Lagrange
equation (\ref{cqg35}) takes the form
\be
\ol x^i_{00}=\{_\la{}^i{}_\nu\}\ol x^\la_0\ol x^\nu_0, \qquad \ol x^0_0=1,
\ee
where $\{_\la{}^i{}_\nu\}$ are the Christoffel symbols of the metric
(\ref{cqg21}) whose components with respect to the coordinates $(x^0,\ol
x^i)$ read
\beq
g_{ij}= -\ol m_{ij}, \qquad g_{0i}=0, \qquad g_{00} =-2f'. \label{cqg51}
\eeq
This metric is Riemannian if $f'>0$ and pseudo-Riemannian if $f'<0$.
 Then the spatial part of the corresponding geodesic
equation
\be
&& \ddot{\ol x}^0 = 0, \qquad \dot{\ol x}^0=1,
\\ 
&&\ddot {\ol x}^i =
\{_\la{}^i{}_\nu\}\dot{\ol x}^\la\dot{\ol x}^\nu
\ee
is exactly the spatial part of the geodesic equation with
respect to the Levi--Civita connection of the metric (\ref{cqg51}) on $TX$.
It follows that, as was  declared above,  the
nonrelativistic dynamic equation (\ref{cqg51}) describes the
nonrelativistic approximation of the geodesic motion in the Riemannian or
pseudo-Riemannian space with the metric (\ref{cqg51}). Note that the spatial
part of this metric is the mass tensor which may be treated as a variable
(Mangiarotti and Sardanashvily, 1998). 

Conversely, let us consider a geodesic motion
\beq
\ddot x^\m=\{_\la{}^\m{}_\nu\}\dot x^\la\dot x^\nu
\label{cqg70}
\eeq
in the presence of a
pseudo-Riemannian metric $g$ on a world manifold $X$. Let $(x^0,\ol x^i)$ be
local hyperbolic coordinates such that $g_{00}=1$, $g_{0i}=0$. This
coordinates define a nonrelativistic reference frame for a local fibration
$X\to\bR$. Then the equation (\ref{cqg70}) has the nonrelativistic limit
\ben
&& \ddot{\ol x}^0 = 0, \qquad \dot{\ol x}^0=1, \nonumber
\\ &&\ddot{\ol x}^i = \{_\la{}^i{}_\nu\}\dot{\ol
x}^\la\dot{\ol x}^\nu
\label{cqg71}
\een
which is the Lagrange equation for the Lagrangian
\be
L=\frac12\ol m_{ij}\ol x^i_0 \ol x^j_0,
\ee
describing a free 
nonrelativistic mechanical system with the mass tensor $\ol m_{ij}=-g_{ij}$.
Relative to another frame $(x^0,x^i(x^0,\ol x^j))$
associated with the same local splitting $X\to\bR$, the nonrelativistic limit
of the equation (\ref{cqg70}) keeps the form (\ref{cqg71}), whereas the
nonrelativistic equation (\ref{cqg71}) is brought into the
Lagrange equation (\ref{cqg47}) for the Lagrangian
\beq
L=\frac12m_{ij}(x^\m)( x^i_0-\G^i)(x^j_0-\G^j). \label{cqg120}
\eeq
This Lagrangian describes a mechanical system in the presence of the inertial
force associated with the reference frame $\G$. The difference between
(\ref{cqg47}) and (\ref{cqg71}) shows that a gravitational force can not model
an inertial force in general. Nevertheless, if the mass tensor in the
Lagrangian
$L$ (\ref{cqg120}) is independent of time, the corresponding Lagrange equation
is a spatial part of the geodesic equation in a pseudo-Riemannian space.

In view of the ambiguity that we have mentioned, the relativization
(\ref{cqg40}) of an arbitrary nonrelativistic quadratic Lagrangian
(\ref{cqg20}) may lead to a confusion. In particular, it can be applied to a
gravitational Lagrangian (\ref{cqg48}) where $f'$ is a gravitational
potential. An arbitrary quadratic dynamic equation can be written in the form
\be
x^i_{00}=-(m^{-1})^{ik}\{_{\la k\m}\} x^\la_0 x^\m_0 + b^i_\m(x^\nu)x^\m_0,
\qquad x^0_0=1,
\ee
where $\{_{\la k\m}\}$ are the Christoffel symbols of some
admissible pseudo-Riemannian metric $g$, whose spatial part is the mass tensor
$(-m_{ik})$, while
\beq
b^i_k(x^\m) x^k_0 + b^i_0(x^\m) \label{cqg61}
\eeq
is an external force. With respect to the coordinates where  $g_{0i}=0$, one
may construct the relativistic equation
\beq
\ddot x^\m= \{_\la{}^\m{}_\nu\}\dot x^\la\dot x^\nu +
\si^\m_\la \dot x^\la, \label{cqg73}
\eeq
where the soldering form $\si$ must fulfill the condition (\ref{cqg46}). It
takes place only if 
\be
g_{ik}b^i_j + g_{ij}b^i_k=0,
\ee
i.e., the external force (\ref{cqg61}) is the Lorentz-type force plus 
some potential one. Then, we have
\be
\si^0_0= 0, \qquad \si^0_k = -g^{00}g_{kj}b^j_0, \qquad \si^j_0=b^j_0.
\ee

The relativization (\ref{cqg73}) exhausts almost all familiar examples.
It means that a wide class of mechanical system can be represented as a
geodesic motion with respect to some affine connection in the spirit of
above mentioned Cartan's idea. 

To complete our exposition,  point  out also
another "relativization" procedure. Let a force
$\xi^i(x^\m)$ in the nonrelativistic dynamic equation (\ref{cqg5}) be a
spatial part of a 4-vector
$\xi^\la$ in  the Minkowski space $(X,\eta)$. Then one can write the
relativistic equation
\beq
\ddot x^\la= \xi^\la -\eta_{\al\bt}\xi^\bt \dot x^\al\dot
x^\la. \label{cqg74}
\eeq
This is the case, e.g., for a relativistic hydrodynamics that we meet usually
in the literature on a gravitation theory. However, this is not a geodesic
equation, and  the nonrelativistic limit $\dot x^0=1$ of the
equation (\ref{cqg74}) does not coincide with the initial nonrelativistic
equation. There are also other variants of relativistic hydrodynamic equations
(Kupershmidt, 1992).

\section{NONRELATIVISTIC JACOBI FIELDS}

Let us consider the quadratic dynamic equation (\ref{cqg100}) and the
equivalent geodesic equation (\ref{cqg11}) with respect to the symmetric
linear connection $\ol K$ (\ref{cqg101}). Its curvature
\be
R_{\la\m}{}^\al{}_\bt =\dr_\la K_\m{}^\al{}_\bt - \dr_\m K_\la{}^\al{}_\bt +
K_\la{}^\g{}_\bt K_\m{}^\al{}_\g - K_\m{}^\g{}_\bt K_\la{}^\al{}_\g
\ee
has the temporal component 
\beq
R_{\la\m}{}^0{}_\bt=0. \label{990}
\eeq
It should be emphasized that our expressions for connections and the
curvature differ in a minus sign from those usually used. Then the equation for
a Jacobi vector field $u$ along a geodesic reads
\beq
\dot x^\bt\dot x^\m(\nabla_\bt(\nabla_\m u^\al) - R_{\la\m}{}^\al{}_\bt u^\la)=
0,\qquad \nabla_\bt\dot x^\al=0, \label{991}
\eeq
where $\nabla$ denote covariant derivatives relative to the
connection $\ol K$ (Kobayashi and Nomizu, 1969). Due to the relation
(\ref{990}), the equation (\ref{991}) for the temporal
component $u^0$ of a Jakobi field takes the form
\be
\dot x^\bt\dot x^\m(\dr_\m\dr_\bt u^0 + K_\m{}^\g{}_\bt \dr_\g u^0)=0.
\ee
We chose its solution 
\beq
u^0=0 \label{993}
\eeq
because all nonrelativistic geodesics obey the constraint $\dot x^0=0$.

In the case of a quadratic Lagrangian $L$ the equation (\ref{991}) coincides
with the Jacobi equation
\be
u^jd_0(\dr_j\dot\dr_iL) + d_0(\dot u^j\dot\dr_i\dot\dr_j L) -u^j\dr_i\dr_jL=0
\ee
for a Jacobi field on solutions of the Lagrange equations for $L$. This
equation is the Lagrange equation for the vertical extension $L_V$ of the
Lagrangian $L$ (Mangiarotti and Sardanashvily, 1998; see also Dittrich and
Reuter, 1992).

Let us consider the quadratic Lagrangian
(\ref{cqg20}) with a Riemannian mass tensor $m_{ij}$. The corresponding
Lagrange equations are equivalent to the geodesic equation (\ref{cqg11}) for
the linear connection
\beq
\ol K_\la{}^0{}_\m=0, \qquad \ol K_\la{}^i{}_\m= (-m^{-1})^{ik}\{_{\la k\m}\},
\label{992}
\eeq
where $\{_{\la k\m}\}$ are the Christoffel symbols of the metric (\ref{cqg21}).
This metric is not necessarily Riemannian. Therefore, given a reference
frame $(x^0,x^\la)$, let us consider another metric
\beq
\ol g_{00}=-1, \qquad \ol g_{0i}=0, \qquad \ol g_{ij}=-m_{ij} \label{994}
\eeq
which is always Riemannian.
However, its covariant derivative with respect to the connection $\ol K$
(\ref{992}) does not vanish. We have
\be
\nabla_\la \ol g_{00}= \nabla_\la \ol g_{ik}=0, \qquad \nabla_\la \ol g_{0k}=
\{_{\la k 0}\}\neq 0.
\ee
Nevertheless, due to the condition (\ref{993}), the well-known formula 
\ben
&& \op\int_a^b(\ol g_{\la\m}\dot x^\al\nabla_\al u^\la \dot x^\bt\nabla_\bt
u^\m + R_{\la\m\al\nu} u^\la u^\bt \dot x^\m\dot x^\nu)
dt + \label{995}\\
&& \qquad \ol g_{\la\m}\dot x^\al\nabla_\al u^\la u'^\m\mid_{t=a} -
\ol g_{\la\m}\dot x^\al\nabla_\al u^\la u'^\m\mid_{t=b} =0 \nonumber
\een
for a
Jacobi vector field $u$ along a geodesic $c$ takes place. Accordingly, the
following assertions also remain true (Kobayashi and Nomizu, 1969).

\begin{prop} \label{1}
If the sectional curvature $R_{\la\m\al\nu} u^\la u^\bt \dot x^\m \dot
x^\nu$   is nonpositive, a geodesic motion has no conjugate points.
\end{prop}

\begin{prop} \label{2}
If the sectional curvature $R_{\la\m\al\nu} u^\la u^\bt v^\m v^\nu$, where
$u$, $v$ are arbitrary unit vectors on a Riemannian manifold $X$ exceeds $k>0$,
then, for every geodesic, the distance between two consecutive conjugate
points is at most $\pi/\sqrt{k}$.  
\end{prop}

For instance, let us consider a one-dimensional motion described by the
Lagrangian
\be
L=\frac12 (\dot x^1)^2 -\f(x^1),
\ee
where $\f$ is a potential. The corresponding Lagrange equations are equivalent
to the geodesic one on the 2-dimensional space $\bR^2$ with respect to the
connection $\ol K$ whose nonzero component is $\{_0{}^1{}_0\}=-\dr_1\f$.
The curvature of $\ol K$ has the nonzero component
\be
R_{10}{}^1{}_0=\dr_1\{_0{}^1{}_0\}=-\dr_1^2\f.
\ee
Choosing the Riemannian metric (\ref{994}) 
\be
\ol g_{11}=-1, \quad \ol g_{01}=0, \quad \ol g_{00}=-1,
\ee
we come to the formula (\ref{995}) 
\be
\op\int_a^b[(\dot x^\m\dr_\m u^1)^2- \dr_1^2\f (u^1)^2]dt=0.
\ee
for a Jacobi vector field $u$ which vanishes at points $a$ and $b$. Then
we obtain from Proposition
\ref{1} that, if $\dr_1^2\f<0$ at points of $c$, this motion has no
conjugate points.
In particular, let us consider the oscillator $\f= k(x^1)^2/2$.  In this case,
the sectional curvature is $R_{0101}=k$, while the half-period of this
oscillator is exactly
$\pi/\sqrt{k}$ in accordance with Proposition \ref{2}.

\end{document}